\documentclass[11pt,openany,openbib]{article}
\usepackage{amssymb}
\usepackage{amsmath}
\usepackage{epic}
\usepackage{eepic}
\usepackage{graphicx}
\DeclareSymbolFontAlphabet{\mathbb}{AMSb}
\usepackage{Vmargin}
\setmargins{1in}{1in}{6.5in}{9in}{0pt}{0pt}{12pt}{42pt}
\usepackage{times}
\usepackage{amsfonts}
\usepackage{eucal}

\newtheorem{theo}{Theorem}
\newtheorem{lm}[theo]{Lemma}

\newtheorem{defn}[theo]{Definition}
\newtheorem{ex}[theo]{Example}

\begin{document}

\title{Diversification in the Internet Economy:The Role of For-Profit Mediators} 
\author{ Sudhir Kumar Singh\\
UCLA \\
suds@ee.ucla.edu 
\and 
Vwani P. Roychowdhury\\
UCLA \& NetSeer Inc.\\
vwani@netseer.com
\and
 Himawan Gunadhi \\
NetSeer Inc. \\
gunadhi@netseer.com
\and
 Behnam A. Rezaei\\
NetSeer Inc.\\
behnam@netseer.com
}
\maketitle

\begin{abstract}  We investigate market
forces that would lead to the emergence of new classes of players in the sponsored search
market. We report a $3$-fold diversification triggered by two
inherent features of the sponsored search market, namely,  \textit{capacity constraints}
and \textit{collusion-vulnerability} of current mechanisms. In the {\em first scenario},
we present a comparative study of two models motivated by capacity constraints - one where
the additional capacity is provided by for-profit agents (or, mediators), who compete for
slots in the original auction, draw traffic, and run their {\em own sub-auctions}, and the
other, where the additional capacity is provided by the auctioneer herself, by essentially
acting  as a mediator and running a {\em single combined auction}. This study was
initiated by us in \cite{SRGR07}, where the mediator-based model was studied. In the
present work, we study the auctioneer-based model and show that this single
combined-auction model seems inferior to the mediator-based model in terms of revenue or
efficiency guarantee due to added capacity. In the {\em second scenario}, we initiate a
game theoretic study of current sponsored search auctions, involving {\em incentive
driven} mediators who exploit the fact that these mechanisms are not collusion-resistant.
In particular, we show that advertisers can improve their payoffs by using the services of
the mediator compared to directly participating in the auction, and that the mediator can
also obtain monetary benefit by managing the advertising burden of its advertisers,
without violating incentive constraints from the advertisers who do not use its services.
We also point out that the auctioneer can not do very much via mechanism design to avoid
such for-profit mediation without losing badly in terms of revenue, and therefore, the
mediators are likely to prevail. Thus, our analysis indicates that there are
\emph{significant opportunities for diversification} in the internet economy and we should
expect it to continue to develop richer structure,
 with room for different types of agents and mechanisms to coexist.
\end{abstract}

\section{Introduction}
\label{intro} Sponsored search advertising is a significant growth market and is
witnessing rapid growth and evolution. The analysis of the underlying models has so far
primarily focused on the scenario, where advertisers/bidders interact directly with a
primary auctioneer or AdNetwork, e.g., they bid for ad-space at leading search engine and
publisher portals. However, the market is already witnessing the spontaneous emergence of
several categories of companies who are trying to mediate or facilitate the auction
process. For example, a number of different AdNetworks have started proliferating, and so
have companies who specialize in reselling ad inventories. Hence, there is a need for
analyzing the impact of such incentive driven and for-profit agents, especially as they
become more sophisticated in playing the game.

Formally, in the current models, there are $K$ slots to be allocated among  $N$ ($\geq K$)
bidders (i.e. the advertisers). A bidder $i$ has a true valuation $v_i$ (known only to the
bidder $i$) for the specific keyword and she bids $b_i$. The expected {\it click through
rate} (CTR) of an ad put by bidder $i$ when allocated slot $j$ has the form $CTR_{i,j}=\gamma_j e_i$
i.e. separable in to a position effect and an advertiser effect. $\gamma_j$'s can be
interpreted as the probability that an ad will be noticed when put in slot $j$ and it is
assumed that $\gamma_j > \gamma_{j+1}$ for all $1 \leq j \leq K$ and $\gamma_j = 0$ for $j > K$.
$e_i$ can be interpreted as the probability that an ad put by
bidder $i$ will be clicked on if noticed and is referred to as the {\it relevance} of bidder
$i$. The
payoff/utility of bidder $i$ when given slot $j$ at a price of $p$ per-click is given by
$e_i\gamma_j (v_i - p)$ and they are assumed to be rational agents trying to maximize
their payoffs.

As of now, Google as well as Yahoo! use schemes closely modeled as RBR(rank by revenue)
with GSP(generalized second pricing). The bidders are ranked in the decreasing order of
$e_ib_i$ and the slots are allocated as per this ranks. For simplicity of notation, assume
that the $i$th bidder is the one allocated slot $i$ according to this ranking rule, then
$i$ is charged an amount equal to $\frac{e_{i+1} b_{i+1}}{e_i}$ per-click. This mechanism
has been extensively studied in recent years\cite{EOS05,Lah06,Var06,LP07,BCPP06,Feng-et}.
The solution concept that is widely adopted to study this auction game is a refinement of
Nash equilibrium independently proposed by Varian\cite{Var06} and Edelman et
al\cite{EOS05}. Under this refinement, the bidders have no incentive to change to another
positions even at the current price paid by the bidders currently at that position.
Edelmen et al \cite{EOS05} calls it  {\it locally envy-free equilibria} and argue that
such an equilibrium arises if agents are raising their bids to increase the payments of
those above them, a practice which is believed to be common in actual keyword auctions.
Varian\cite{Var06} called it {\it symmetric Nash equilibria(SNE)} and provided some
empirical evidence that the Google bid data agrees well with the SNE bid profile. In
particular, an {\bf SNE} bid profile $b_i$'s satisfy
\begin{align}
\label{sne}
(\gamma_i - \gamma_{i+1}) v_{i+1} e_{i+1} + \gamma_{i+1} e_{i+2} b_{i+2}  \leq \gamma_{i} e_{i+1} b_{i+1} \nonumber \\
 \leq (\gamma_i - \gamma_{i+1}) v_{i} e_{i} + \gamma_{i+1} e_{i+2} b_{i+2}
\end{align}
for all $ i=1,2,\dots, N$.
Now, recall that in the RBR with GSP mechanism, the bidder $i$ pays an amount $\frac{ e_{i+1} b_{i+1} }{e_i}$ per-click,
therefore the expected payment $i$ makes per-impression is
$\gamma_i e_i \frac{ e_{i+1} b_{i+1} }{e_i} = \gamma_i e_{i+1} b_{i+1}$.
Thus the best {\bf SNE} bid profile for advertisers (worst for the auctioneer) is minimum bid profile
possible according to Equation \ref{sne} and is given by
\begin{equation}
 \gamma_{i} e_{i+1} b_{i+1} = \sum_{j=i}^K (\gamma_j - \gamma_{j+1}) v_{j+1} e_{j+1}
\end{equation}
and therefore, the revenue of the auctioneer at this minimum {\em SNE} is
\begin{equation}
 \sum_{i=1}^K \gamma_{i} e_{i+1} b_{i+1} = \sum_{i=1}^K \sum_{j=i}^K (\gamma_j - \gamma_{j+1}) v_{j+1} e_{j+1}= \sum_{i=1}^K  (\gamma_j - \gamma_{j+1}) j v_{j+1} e_{j+1}.
\end{equation}
\\

In the present work, we continue to pursue a perspective transversal to the previous
studies such as in the above mentioned articles, initiated in \cite{SRGR07} by us. We
investigate the emergence of scenarios, where advertisers/bidders do not interact directly
with the auctioneers, as well as, the elements that trigger such a diversification in the
sponsored search market. We report a $3$-fold diversification in the terms of
\begin{itemize}
 \item the emergence of new market mechanisms
\item  the emergence of new for-profit agents, and
\item  the participation of a wider pool of bidders/advertisers.
\end{itemize}
Such a diversification is triggered by two inherent features of the sponsored search market:
\begin{itemize}
\item  {\em Capacity constraints:} One natural constraint in the sponsored search
  advertising framework arises from the fact that there is a
 limit on the number of available slots (or slots that receive any clicks from users),
 especially for the popular keywords, and as a result,
 a significant pool of advertisers are left out.
 In  \cite{SRGR07} we intitiated a study of the emergence of diversification in the
sponsored search market triggered by such capacity constraints in the sense
  that new market mechanisms, as well as, new for-profit agents are likely to emerge to combat or to make profit from
the opportunities created by shortages in ad-space inventory.
In  \cite{SRGR07}, we proposed a model where the additional capacity is
 provided by for-profit agents (or, mediators), who compete for slots in the original auction, draw traffic, and
run their own sub-auctions. It was shown that the revenue of the auctioneer, as well
as the social value (i.e. efficiency), always increase when mediators are involved. An
important question left open in that paper was - what if the auctioneer wants to provide
the additional capacity herself by essentially acting herself as a mediator and running a
single {\em combined auction}? Does the revenue of the auctioneer/efficiency improves or
does it degrade in such a model? We investigate this direction in Section \ref{capacity}
of the present work. Following \cite{SRGR07}, the quality of the additional capacity is
measured by a {\it fitness} factor. We show that, unlike the mediator-based model,  often
there is a tradeoff between the revenue and the capacity, and there is a phase transition
from possibly a gain in terms of revenue to a loss as the {\it fitness} increases, meaning
that there is a critical fitness value beyond which the auctioneer always loses in
revenue. However, there exist scenarios where the revenue of the auctioneer could indeed
increase by increasing capacity. In the case of efficiency, the result is more in
consonance with the mediator-based model, i.e., the efficiency increases as fitness
increases. However, unlike the mediator-based model, the efficiency could indeed decrease
by increasing capacity. Nevertheless, there is a phase transition and a critical fitness
beyond which efficiency is never lost by adding capacity.

\item  {\em Collusion-vulnerability of current mechanisms:}
In Section \ref{fpmed}, we initiate a game theoretic study of current sponsored search
auctions, involving {\em incentive driven} mediators who exploit the fact that these
mechanims are not collusion-resistant. As we know, the market is already witnessing the
spontaneous emergence of several categories of companies who are trying to mediate or
facilitate the auction process and it is likely that such mediators would exploit such
opportunities. In particular, we will show that advertisers can improve their payoffs  by
using the services of the mediator compared to directly participating in the auction and
mediator can also obtain monetary benefit by managing the advertising burden of its
advertisers and in fact at the same time being compatible with incentive constraints from
the advertisers who do not use its service. We also point out that the auctioneer can not
do very much via mechanism design to avoid such for-profit mediation without losing badly
in terms of revenue, and therefore, the mediators are likely to prevail. Thus, our results
show that mediators can play a significant role in sponsored search auctions, and can
potentially impact the revenues earned by the auctioneer.

\end{itemize}
\section{Diversification triggered by capacity constraints}
\label{capacity}
In this section, we study the emergence of diversification in the sponsored search market triggered by capacity constraints in the sense
 that new market mechanisms, as well as, new for-profit agents are likely to emerge to combat or to make profit from
the opportunities created by shortages in ad-space inventory.
Such a study was initiated by us in \cite{SRGR07}. In  \cite{SRGR07}, we proposed a model where the additional capacity is
 provided by for-profit agents (or, mediators), who compete for slots in the original auction, draw traffic, and
run their own sub-auctions. It was shown that the
the revenue of the auctioneer,
as well as the social value (i.e. efficiency), always increase when mediators are involved.
As we pointed out there, internet economy is already evolving to such a diversification
\footnote{For example, the keyword ``personal loans'' or ``easy loans'' and the mediator ``personalloans.com''.}.
An important question left open in that paper was - what if the auctioneer wants to provide the additional capacity
herself by essentially acting herself as a mediator and running a combined auction?
Does the revenue of the auctioneer/efficiency improves or does it degrade in such a model?
At a philosophical level we can interpret the comparative study of the two different
models for providing additional capacity- one via for-profit agents and other via auctioneer herself as
the study of {\it evolution versus design} in the case of sponsored search market.
For a detailed description of the model where the additional capacity is provided a for-profit agent(mediator),
the readers are referred to \cite{SRGR07} wherein the following theorem is proved.

\begin{theo}
\label{fpmcap}
Increasing the capacity via mediator improves the revenue of the auctioneer, as well as, the efficiency.
\end{theo}

\subsection{The model where additional capacity is provided by the auctioneer}
\label{modcap}
Now we formally describe the model where the additional capacity (i.e. additional slots) is provided by the same auctioneer/search engine.

\begin{itemize}

\item {\bf Additional slots:}
The additional slots are obtained by forking {\em one} of the original slots. By forking
we mean the following: the auctioneer puts her own ad/link  in that slot, and on the
associated landing page, she puts some information relevant to the specific keyword along
with space for additional slots. We have considered the single fork case for the sake of
simplicity of presentation and so that the calculations do not get unwieldy, but the
results can be easily extended to the case where the auctioneer forks multiple slots and
adds additional capacity.

\item {\bf Properties of the additional slots and {\it fitness} factor:} The
quality of the additional slots is measured by a {\it fitness factor}. Let the probability
associated with the ad put by the auctioneer for creating additional capacity to be
clicked, if noticed,  be denoted as $\tilde{f}$. Moreover, the position-based CTRs for the
additional slots in the landing page will in general be different than on the main page,
and it might actually improve, say by a factor of $\alpha$. This means that the position
based CTR for the $j$th additional slot on the associated landing page is modeled as
$\alpha \gamma_j$. Therefore, we can define a fitness factor $f$ to indicate the effective
quality of additional slots being created, which is equal to $\tilde{f}\alpha$. Thus, if
the original slot being forked is $l$, and there are $L$ additional slots being created on
the landing page, then the {\it effective} position based CTRs for the additional slots
thus obtained are $\gamma_{l} f \gamma_1, \gamma_{l} f \gamma_2, \dots, \gamma_{l_i} f
\gamma_{L}$ respectively. Clearly, $ f \gamma_1 < 1$; however, $f$ itself could be greater
than $1$.

\item {\bf A single combined auction:} Auctioneer runs a single combined auction
to sell original slots together with the additional ones.
Therefore, a bidder is allowed {\em only} to bid for all slots (original slots plus the
additonal ones) together and not for the two kind of slots individually. This is unlike
the case with the additional capacity  being created by a for-profit agent (mediator)
\cite{SRGR07} where the mediator runs her own sub-auction.  Further, for the comparative
analysis in the following, we assume that this combined auction is run via RBR with GSP
i.e. the mechanism currently being used by Google and Yahoo! and the solution concept we
use is Symmetric Nash Equilibria(SNE)/locally envy-free equilibria\cite{EOS05,Var06}.

\item {\bf New position based CTRs:} Note that in the combined auction there
are now $\tilde{K}=K+L-1$ slots and for each slot there will be a probability of being
noticed if an advertiser is assigned to that slot i.e. its position based CTR. We rename
the slots in the decreasing order of their CTRs. That is, the $j$th slot is the one having
$j$th maximum of the elements from the set $\{\gamma_1, \gamma_2, \dots,
\gamma_{l-1},\gamma_{l+1}, \dots, \gamma_K\} \cup \{\gamma_l f \gamma_1, \gamma_l f
\gamma_2, \dots, \gamma_l f \gamma_{L}\}$ and its CTR is denoted by $\tilde{\gamma}_j$.
For the sake of simplicity, we assume that there are no ties i.e. no two slots have the
same position based CTRs. Therefore, like $\gamma_j$'s we have $\tilde{\gamma}_j
>\tilde{\gamma}_{j+1}$ for all $ 1 \leq j \leq K+L-1$ and $\tilde{\gamma}_j=0$ for all $j
\geq K+L$. Further note that,  $\tilde{\gamma}_j = \gamma_{j}$ for $ j \leq l-1$, and
$\tilde{\gamma}_l < \gamma_{l}$. Therefore
 $\tilde{\gamma}_j -\tilde{\gamma}_{j+1} = \gamma_j - \gamma_{j+1}$ for $ j < l-1$,
$\tilde{\gamma}_{l-1} -\tilde{\gamma}_l > \gamma_{l-1} - \gamma_l$, and  $\tilde{\gamma}_j -\tilde{\gamma}_{j+1}$ could be
greater than or less than $\gamma_j - \gamma_{j+1}$ for $ l \leq j \leq K $ depending on how the new position based CTRs are distributed among
the old ones.

\item {\bf Capacity:} The {\it capacity} is defined as the sum of position based CTRs. Thus the capacity in the original model without
the additional slots  is $C_0=\sum_{j=1}^K \gamma_j$ and in the model with the additional slots it is
$C=\sum_{j=1}^{K+L-1} \tilde{\gamma}_j = \sum_{i=1, i\neq l}^K \gamma_j + \gamma_l f \sum_{i=1}^L \gamma_i$.
Note that for a fixed $L,l$, the capacity increase iff $f$ increases, for a fixed $l,f$, it
increase iff $L$ increases.

\item{\bf Notation:} The bidder/advertiser $i$'s value and bid are denoted by $v_i$ and $b_i$ respectively
and her relevance (i.e. quality score) is denoted by $e_i$. Further, for simplicity we will denote $e_iv_i$ by
$s_i$ and $e_i b_i$ by $r_i$.
\end{itemize}

\subsection{Revenue of the auctioneer}
Now we discuss the loss or gain in the auctioneer's revenue due to the added capacity as
per the model in Section \ref{modcap}. First, recall that the  auctioneer's revenue always
increases in the model where the additional capacity is provided by a mediator (Theorem
\ref{fpmcap}) and in fact the revenue of the auctioneer increases as the {\it fitness} of
the mediator increases, thus there is no conflict between the revenue and the capacity in
that model. However, as we shall see in the following, in the model of Section
\ref{modcap} often there is a tradeoff between the revenue and the capacity.   First, we
define the {\it value of capacity} and then establish a sufficient condition for the gain
in the revenue of the auctioneer in the following theorem whose simple proof is deferred
to the Appendix.

\begin{defn}
\label{voc} {\bf Value of capacity:}  Let $R_0$ be the original revenue of the auctioneer
without added capacity and $R$ be the new revenue of the auctioneer after adding capacity
at their corresponding minimum {\em SNE} \cite{EOS05,Var06}, then the  ``value of
capacity'' is defined as $\frac{R-R_0}{R_0}$ i.e. the relative gain in the revenue of
auctioneer per impression.
\end{defn}

\begin{theo}
\label{rev1}
For a given L, if  $\exists l \leq K$ such that
\begin{align}
\label{eta}
  \eta > 1- \left(\frac{(\gamma_l - \tilde{\gamma}_l)(l-1) s_l + \sum_{j=K+1}^{K+L-1}  (\tilde{\gamma}_j - \tilde{\gamma}_{j+1}) j s_{j+1} }{\sum_{j=l}^K (\gamma_j - \gamma_{j+1}) j s_{j+1}}\right)\\
\textrm{  where   }  \eta =\min_{K \geq j \geq l} \frac{\tilde{\gamma}_j -\tilde{\gamma}_{j+1}}{\gamma_j - \gamma_{j+1}}
\end{align}
then the value of capacity is positive, i.e., revenue of the auctioneer increases by
adding capacity.
\end{theo}
We first provide an example to confirm that the above theorem does not give a vacuous sufficient condition and the {\it value of capacity} can indeed be positive.
\begin{ex}
\label{example1}
Let $l=K$ and geometrically decreasing $\gamma_j$'s\cite{AG07,Feng-et} i.e. $\gamma_j = r^{j-1}, 1 \leq j \leq K$ for some $r <1$ and $0$ otherwise.
Then $\tilde{\gamma}_j = r^{j-1}$ for $ 1 \leq j \leq K-1$ and $\tilde{\gamma}_j = f r^{j-1}$ for $K \leq j \leq K+L-1$ and $0$ otherwise.
Also, let $j s_{j+1} \geq (j-1) s_j$ for all $ K+1 \leq j \leq K+L-1$ and $(K-1) s_K > K s_{K+1}$. Then, the condition for Theorem \ref{rev1} is satisfied.
Detailed calculations are provided in the Appendix.
\end{ex}

Now having shown that the  {\it value of capacity} could be positive, we would like to see
how much worse it could be, and a clearer tradeoff between revenue and {\it fitness}, if
any. The following lemmas formalize some worst case scenarios. The proofs are moved to the
Appendix.

\begin{lm}
\label{l1revloss}
Let $l=1$ and $\gamma_j$'s and $s_i$'s satisfy $(\gamma_1 - \gamma_2) \geq (\gamma_j - \gamma_{j+1})$ for all $ 1 \leq j \leq K$ and
$ (j-1) s_j \geq j s_{j+1}$ for all $ j \geq 2$, then there exists no fitness factor $f$ such that the value of capacity is positive.
\end{lm}

\begin{lm}
\label{l2revloss}
Let $s_i$'s satisfy $ (j-1) s_j \geq j s_{j+1}$ for all $ j \geq 2$, then for any $l \geq 2$, the value of capacity is a decreasing function of $f$ and $L$.
\end{lm}

Recall that the bidders are characterized by their true valuations $v_i$'s and their
relevance scores $e_i$'s and $s_i=e_iv_i$. Thus, there is a wide pool of bidders
satisfying the conditions in the above lemmas, and therefore indicating a significant
tradeoff between revenue of the auctioneer and the capacity. Intuitively, the conditions $
(j-1) s_j \geq j s_{j+1}$ in Lemma~\ref{l2revloss} say  that the $s_i$'s are well
separated, and therefore the payments that the bidders make at {\bf SNE} are also well
separated. Increasing capacity (via increasing $f$ or $L$) means essentially selling a
fraction of the clicks at a lower price. When $s_i$'s are well separated, the extra
revenue coming from the newly accommodated bidders still fall short of  that lost due to
lower payments from the other bidders. Further, the Lemma \ref{l2revloss} suggests that
there is a phase transition from possibly positive value of capacity to negative as  $f$
increases, and there is a critical $f$ beyond which the auctioneer always loses.

\subsection{Efficiency}
Now let us look at the change in efficiency due to added capacity as per the model in Section \ref{modcap}.
First, recall that the efficiency always increases in the model where the additional capacity is provided by a mediator
(Theorem \ref{fpmcap}) and in fact
the efficiency increases as the {\it fitness} of the mediator increases, thus there is no conflict between the efficiency and the
capacity in that model.
As we will show in the following, even though the efficiency might decrease sometimes by adding capacity via the auctioneer
controlled model, unlike the case of revenue of the auctioneer where there is a decrease whenever fitness increases,
the efficiency will always increase when fitness increases. Thus there is a phase transition and a critical fitness
beyond which efficiency is never lost by adding capacity.
First, we establish a sufficient condition for the gain in the efficiency
in the following theorem whose simple proof is deferred to the Appendix.

\begin{theo}
\label{eff1}
For a given L, if $\exists l \leq K$ such that
\begin{align}
\beta > 1- \frac{ \sum_{j=K+1}^{K+L-1} \tilde{\gamma}_j s_j}{\sum_{j=l}^K \gamma_j s_j} \\
\textrm{   where   } \beta = \min_{K \geq j \geq l} \frac{\tilde{\gamma}_j}{\gamma_j},
\end{align}
then the efficiency improves.
\end{theo}

We first provide an example to confirm that the above theorem does not give a vacuous sufficient condition and
the efficiency can indeed improve.
\begin{ex}
\label{example2}
Let $l=K$ and geometrically decreasing $\gamma_j$'s as in Example \ref{example1} and let $s_i$'s satisfy $s_{K+j}= \alpha^j s_K$
for some $\alpha <1$.
Then, the condition for Theorem \ref{eff1} is satisfied when $f > \left(\frac{1-\alpha r }{1- \alpha^L r^L}\right)$.
Detailed calculations are provided in the Appendix.
\end{ex}

Now in the following theorem we will show that the efficiency always increases when fitness increases
and there is a phase transition and a critical fitness
beyond which efficiency is never lost by adding capacity via model in Section \ref{modcap}.
Again, we move the proof to the Appendix.

\begin{theo}
\label{eff2}
The efficiency is an increasing function of fitness $f$.
\end{theo}

\section{Diversification triggered by collusion - vulnerabilty of  mechanims}
\label{fpmed}
A mediator is a well-known construct in game theory, and is an entity that plays on behalf
of some of the agents who choose to use its services, while the rest of the agents
participate in the game directly. In this section, we initiate a game theoretic study of sponsored search
auctions, such as those used by Google and Yahoo!, involving {\em incentive driven}
mediators. We refer to such mediators as {\em for-profit} mediators, so as to distinguish
them from  mediators introduced in prior work\cite{med}, who have no monetary incentives, and are
driven by the altruistic goal of implementing certain desired outcomes despite the financial cost
incurred. As we know, the marketplace is mostly about incentives- a game between selfish
agents- and it would be interesting to study mediators which are not altruistic.
In particular, we will show that advertisers can improve their payoffs by using the services of the mediator
compared to directly participating in the auction and mediator can also obtain monetary
benefit by managing the advertising burden of its advertisers and in fact at the same time
being compatible with incentive constraints from the advertisers who do not use its service.
The simple
intuition behind the above result comes from the observation that since the mediator has
more information about and more control over the bid profile
than any individual advertiser, she could possibly
modify their bids, before reporting to the auctioneer (search engine), in a manner to
improve their payoff and could retain a fraction of the improved payoff.
However, the above intuition is of course not a formal game-theoretic argument
why the collusion via mediation will work as we do also need to argue that the other
advertisers (the advertisers who do not bid via the mediator) still do not have
incentives to change their  bids and slot positions.  In the following sub-sections we present a game theoretic
analysis for the sponsored search auctions via mediation and show that the intuition given above
is indeed true.
We consider the case where there is only one mediator and the analysis in
the other cases essentially remains similar.

\subsection{Definitions and Model Setup}

\begin{itemize}

\item {\bf For-profit mediator:}  A mediator is a reliable entity, which can play on the behalf of agents in a given game,
however it can not enforce the use of its services, and each agent is free to participate in the game directly. In our study,
a {\it for-profit}
mediator is a mediator who is a selfish agent trying to maximize a well defined payoff while bidding on the behalf
of the advertisers/bidders who choose to use its services. Throughout the Section \ref{fpmed}, we will simply use the word {\it mediator}
to refer to the {\it for-profit} mediator.

\item {\bf $M$-bidders:} The advertisers/bidders who bid via the mediator will
be called {\it $M$-bidders}. These bidders report their bids to the mediator. As usual, these bidders are selfish agents and
they will stick to using the services of the mediator as long as it provides them a better payoff than what they could have obtained
without using its service. Let us denote the set of $M$-bidders by $M$.

\item {\bf $I$-bidders:} The advertisers who choose not to bid via the mediator will
be called {\it $I$-bidders}. These bidders report their bids to the auctioneer directly.
Let us denote the set of $I$-bidders by $I$.

\item {\bf Incentive constraints:}
\begin{itemize}

\item The bids $b_1, b_2, \dots, b_L,b_{L+1}, \dots b_K \dots $ that the
bidders report, either to mediator or to the auctioneer, are at the equilibrium  of the
auction without any mediation under a proper solution concept. Therefore, in the original
game the bidders have no incentives to defect from their current positions\footnote{It is
reasonable to assume this as the auction process has been going for a while now. Further,
this requirement can be relaxed -- the mediator first bids on the behalf of the M-bidders
to figure out and evolve to an equilibrium before implementing her strategy to modify
their bids.}. For the Google and Yahoo! like auctions, the solution concept we use is Nash
equilibria and its refinement symmetric Nash equilibria (SNE). We also consider truthful
mechanisms.

\item  $M$-bidders do not want to change the positions they get via directly
reporting to the auctioneer at equilibrium
of the auction without any mediation \footnote{Relaxing this condition gives  more freedom
to the mediator and she could possibly do even better by changing their positions as
illustrated later in the paper, however advertisers might not like to go down in slot
position due to decrease in traffic as well as branding impression value.}. However, they
give the mediator the right to change their bids before reporting to the auctioneer for a
potential increase in their payoffs.
\end{itemize}

\item {\bf Payoff of the mediator:} The  mediator's
payoff is defined to be a fixed fraction of the total improvement in payoffs from the $M$-bidders
over what they could have obtained without using her service. Formally, let $p_i$ denote the amount
bidder $i \in M$ pays {\it per-click} in the game without mediator and $\tilde{p}_i$ denotes the amount the mediator pays {\it per-click}
on her behalf. Also let $\sigma(i)$ be the slot position assigned to the bidder $i$, then
the payoff of the mediator denoted by $U_M$ is defined as $\alpha \sum_{i \in M} CTR_{i,\sigma(i)} (p_i -\tilde{p}_i)$ for a fixed
$0 < \alpha < 1$.  Further, whenever an advertiser $i \in M$ gets a click, an amount equal to $(1-\alpha) \frac{(p_i -\tilde{p}_i)}{|M|}$
is given to each one of the $M$-bidders, and the mediator keeps an amount equal to $\alpha(p_i -\tilde{p}_i)$ for herself.
Therefore, as long as the payoff of the mediator is positive,  no $M$-bidder will have
an incentive to leave the services of the mediator. Moreover, we would like to mention that for the purpose of incentive analysis,
it is enough that the mediator provides a positive extra amount to the $M$-bidders whenever she obtains a positive payoff
and it does not actually matter how she allocates it amongst the $M$-bidders. This is in keeping with the
coalitional game with transferable payoffs\cite{ORgamebook,GH05}.

\end{itemize}

\subsection{Designing for-profit mediators}

Let us first consider the RBR scheme with GSP
currently being used by Google and Yahoo!. The advertisers are ranked according to $r_i =
e_i b_i$ where $e_i$ is the relevance (quality score) of the bidder $i$. Let us rename the advertisers by this
ranking i.e. $r_1 > r_2 > \dots > r_L > r_{L+1}> \dots > r_K >r_K > \dots > r_N$,
therefore the $i$th bidders pays $\frac{e_{i+1}b_{i+1}}{e_i} = \frac{r_{i+1}}{e_i}$ per-click under GSP.
The solution concept we adopt is Nash equilibrium and its refinement {\em SNE}. Then we shall move to truthful mechanisms.

{\bf At Symmetric Nash Equilibrium:}
Let us first analyze the incentive and revenue properties for the case where the top $L$ advertisers are the $M$-bidders
i.e. $M = \{1,2,\dots,L\}$.
The bids $b_1, b_2, \dots, b_L,b_{L+1}, \dots b_K \dots b_N$ are at the SNE of the auction without any mediation,
therefore in the original game the bidders have no incentives to defect from their current positions.
Now the problem is that how should the mediator modify the bids of $M$-bidders to improve her payoff while maintaining the
incentives constraints from the $I$-bidders. In particular, can the mediator modifies the bids of  $M$-bidders so that no
$I$-bidder $i$ has an incentive to change her bid  $b_i$, as well as, the mediator's payoff is
positive\footnote{Note that we are not necessarily looking for the best possible strategy of the mediator so as to maximize her payoff but
merely the strategies which can give her a positive payoff.}?
Let the mediator modify the bids such that $r_i^{'}=r$ for $i=1,2,\dots,l$ and $r_i^{'}=r_i$ for $i=l+1, \dots, L$, then what $r$ and $l$
should she choose\footnote{The bids will actually be modified so that $r_i^{'}=r + (L-i) \epsilon$ for an infinitesimally small $\epsilon > 0$.
In practice, this $\epsilon$ can not be less than $\$0.01$, however for the purpose of analysis, as in earlier works,
we assume that it is a continuous parameter that
can be made infinitesimally small.}. The auctioneer sees the bid profile
$r \geq r \geq \dots \geq r > r_{l+1} > \dots >r_L > r_{L+1} > \dots > r_K > \dots $.
Since the original bid profile was at SNE, no $j \geq L+1$ would like to deviate to any other position $s$ for $l \leq s \leq K+1$.
Now, only position a $ j \geq L+1$ can deviate to is the position $1$ (she can not do better than this by moving to $2, \dots, l-1$
for she will be paying same price to get less clicks).

The condition that the $I$-bidders do not want to move to position $1$ is
\begin{displaymath}
\begin{array}{ll}
\gamma_j e_j(v_j - \frac{r_{j+1}}{e_j}) \geq \gamma_1 e_j (v_j - \frac{r}{e_j}) &  \forall j \geq L+1 \\
\gamma_j (e_jv_j - r_{j+1}) \geq \gamma_1 (e_j v_j - r) &  \forall j \geq L+1 \\
\therefore  r \geq (1 - \frac{\gamma_j}{\gamma_1}) e_j v_j + \frac{\gamma_j}{\gamma_1} r_{j+1} &  \forall j \geq L+1
\end{array}
\end{displaymath}
\begin{displaymath}
\textrm{Let    }  r^{*} = \max_{j \geq L+1} \{(1 - \frac{\gamma_j}{\gamma_1}) e_j v_j + \frac{\gamma_j}{\gamma_1} r_{j+1}\}
\end{displaymath}
then any selection of $r$ such that $ r \geq r^{*}$ and $r> r_{l+1}$ is fine at SNE and the mediator chooses
$r=r^{*}$ and an $l$ such that $ r_l \geq r > r_{l+1}$. Note that such an $l \geq 2$ always exists as $ r \leq r_2$
(for the $I$-bidders did not want to move to position $1$ in the original game at SNE)\footnote{Of course, the mediator will not be able
to choose such a $r$ and $l$ all at once and will rather evolve to it by trying suitable $r_l$'s.},\footnote{Sometimes for example when
$l=2$ in the above, the mediator could possibly do even better by modifying bids as
$ r_1 > r_2 > r \geq r \geq \dots \geq r > r_{l+1} > \dots >r_L > r_{L+1} > \dots > r_K > \dots $ and so on.}.
The mediator now pays an amount $\frac{r}{e_i}$ {\it per-click} on the behalf of the bidder $i \in M$ and
therefore the total expected payment {\it per-impression} that the mediator makes on the behalf of $M$-bidders is
$(\sum_{j=1}^{l-1} \gamma_j )r + \sum_{j=l}^{L} \gamma_j r_{j+1}$ and the payoff of the mediator is
\begin{displaymath}
U_M = \alpha \sum_{j=1}^{l-1} ( r_{j+1} - r) \gamma_j
\end{displaymath}
and each $M$-bidder gets an expected extra payoff of $ (1- \alpha) \frac{1}{L} \sum_{j=1}^{l-1} ( r_{j+1} - r) \gamma_j$ {\it per-impression}
compared to what they would get in the auction without mediation.
It is clear that this is at the expense of the loss in the revenue of the auctioneer.
Note that in the case when there are only $M$-bidders and no $I$-bidders, the
auctioneer gets the minimum price set for all the slots. Now let us illustrate
the above analysis by an example listed in Table \ref{med1} wherein
the bid profile $\{r_i\}$ is first verified to be at symmetric Nash equlibrium in Table \ref{med2} by recalling that
to verify this we need only check the equilibrium condition for one slot up and one slot down positions
(locally envy free property)  and finally a suitable $r$ is chosen in Table \ref{med3}.
All the tables are moved to the appendix.

Now let us consider the case when the $M$-bidders are not necessarily the top ones but
$ l+1, l+2, \dots, l+L$. The mediator modifies the bids such that
$r_j^{'}=r$ for $j=l+2, \dots, l+s-1$ for some $s \leq L$ and $r_j^{'}=r_j$ otherwise, therefore
the auctioneer sees the bid profile
$r_1 > r_2 > \dots > r_l > r_{l+1} > r \geq r \dots \geq r > r_{l+s} > \dots > r_{l+L} > r_{l+L+1} > \dots $.

As in the earlier analysis, at SNE, the only condition that need to be checked is that no $j \leq l$ or $j \geq l+L+1$
would want to deviate to the $(l+1)$th position. Therefore, we must have for  $j \leq l$ and  $j \geq l+L+1$ ,
\begin{displaymath}
\begin{array}{l}
\gamma_j e_j(v_j - \frac{r_{j+1}}{e_j}) \geq \gamma_{l+1} e_j (v_j - \frac{r}{e_j})  \\
\gamma_j (e_jv_j - r_{j+1}) \geq \gamma_{l+1} (e_j v_j - r) \\
\therefore  r \geq (1 - \frac{\gamma_j}{\gamma_{l+1}}) e_j v_j + \frac{\gamma_j}{\gamma_{l+1}} r_{j+1} .
\end{array}
\end{displaymath}
\begin{displaymath}
\textrm{Let    }  r^{*} = \max_{\{j \leq l\} \cup \{ j \geq l+L+1 \}} \{(1 - \frac{\gamma_j}{\gamma_{l+1}}) e_j v_j + \frac{\gamma_j}{\gamma_{l+1}} r_{j+1}\}
\end{displaymath}
then clearly $r^{*} \leq r_{l+2}$ and choosing any $ r \geq r^{*}$ and an $s$ such that $r_{l+s-1} \geq r > r_{l+s}$ is fine at SNE.
However, in this case or in the case when $M$-bidders are the top ones, such a $r$ to improve their payoffs might not always exist
as can been seen by considering the example mentioned above when $M = \{2,3,4,5\}$.
However, mediator could possibly improve even in these cases, if the $M$-bidders do not mind moving up in positions,
as she does not neccesarily have to satisfy the incentive constraints from higher position $I$-bidders to not change their current
positions.

Similar analysis holds when there are different groups of $M$-bidders such that all advertisers in a group bid
via the same mediator whereas different groups may have different mediators associated with them.
In fact, in this case the maximization will be over smaller sets and the mediators could possibly do even better.

A possibility not analyzed above is that whether the mediator can do better by moving the positions of
the advertisers either individually or sliding them all together. Consider the example given earlier and
let mediator slide every $M$-bidder one slot down by modifying the bid profile so that
$r_i^{'}=12$ for all M-bidders as shown in the Table \ref{med4}. It can be verified as
before that it is still at SNE and in fact in this case mediator's payment on behalf of $M$-bidders is much lesser
compared to the earlier case. This suggests that indeed the mediator could do better by
moving slot positions. However, advertisers might not like to change positions, at least not to the lower slots due to
associated branding impression values coming from higher slots and even though their payoff might increase by allowing so,
they might not like to lose in terms of traffic which decreases by going down. \\

{\bf At non-Symmetric Nash Equilibrium:}
If the bid profile is at non-symmetric Nash equilibrium an analysis similar to the case of {\em SNE} can be adopted.
For example, when $M=\{1,2,\dots,L\}$ the mediator modifies the bids as
 $r_i^{'}=r$ for $i=2,\dots,l$ and $r_i^{'}=r_i$ for $i=1, l+1, \dots, L$ and the condition on $r$ now is
\begin{displaymath}
 r \geq \max_{j \geq L+1} \{(1 - \frac{\gamma_j}{\gamma_2}) e_j v_j + \frac{\gamma_j}{\gamma_2} r_{j+1}\}.
\end{displaymath}
\\

{\bf Truthful mechanisms:}
Truthful mechanisms are considered to be very desirable from the advertisers' perspective
since truth-telling is a dominant strategy for every one and the advertises do not need
to be sophisticated to play the auction game. However, as we argue below it is
more vulnerable to for-profit mediation and even the mediators
need not be sophisticated in this case, unlike the ones discussed earlier in the paper.
In regard to position auctions, Aggarwal et al.\cite{AGM06} presented a truthful
mechanism called {\it laddered auction},
which is compatible with a given weighted ranking function such as RBR, and is the unique truthful auction
given this ranking function.
Now, if the mediator modifies the bids of the $M$-bidders in a manner so that their slot
positions (i.e. ranks) do not change, the $I$-bidders still report truthfully as its
a dominant strategy. The mediator could choose such a minimum possible
bid profile to get the best improvement in payoffs and in particular modifying
every $M$-bidder's bid to a value only infinitesimally more than just enough to retain the position of the least ranked $M$-bidder
suffices. \\

Note that in above construction of for-profit mediators,
we were not necessarily looking for the best possible strategy of the mediator so as to maximize her payoff but
merely the strategies which could give her a positive payoff and the $M$-bidders were the ones assigned to the concecutive slots.
More sophisticated
mediators, such as where $M$-bidders need not be concecutive, should also be investigated.
In general, if the mediator is sophisticated enough to exploit her best
possible strategy, how does the modified bid profile look at equilibrium?
We leave these directions open for future studies. \\

{\bf Collusion-proof mechanisms:}
Having given some constructions of for-profit mediators, it is natural to ask whether
there exist mechanisms
which are impervious to collusion via for-profit mediation and if they do
how do they effect the revenue of the auctioneer i.e. can we design collusion-resistant mechanisms with
good revenue guarantee?
The answer is in negative and
the results of Golderbeg and Hartline\cite{GH05} can be adapted to the case of sponsored search auctions
to show that only mechanisms which are imprevious to collusion are posted-price i.e. that post a ``take it or leave it'' price
for each advertiser, and the revenue guarentee from such mechanism could indeed be very bad in the worst case.
However, using consensus-estimate technique of \cite{GH03}, the authors in \cite{GH05}
design randomized collusion-resistant mechanisms if one asks
incentive constraints to be maintained only with high-probability.
Nevertheless, as these results are mostly asymptotic in nature, they do not seem
applicable to the case of sponsored search auctions, due to the smaller number of slots.
Thus the auctioneer can not do very much via mechanism design to avoid for-profit mediation
and the mediators are likely to prevail.

\section{Concluding remarks}
In the present work, we investigated market
forces that would lead to the emergence of new classes of players in the sponsored search
market. We reported a $3$-fold diversification,
 in the terms of the emergence of new market mechanisms, the emergence of new for-profit agents, and
the participation of a wider pool of bidders/advertisers,
triggered by two
inherent features of the sponsored search market, namely,  \textit{capacity constraints}
and \textit{collusion-vulnerability} of current mechanisms. 

In the {\em first scenario},
we present a comparative study of two models motivated by capacity constraints - one where
the additional capacity is provided by for-profit agents (or, mediators), who compete for
slots in the original auction, draw traffic, and run their {\em own sub-auctions}, and the
other, where the additional capacity is provided by the auctioneer herself, by essentially
acting  as a mediator and running a {\em single combined auction}. This study was
initiated by us in \cite{SRGR07}, where the mediator-based model was studied. In the
present work, we study the auctioneer-based model and show that this single
combined-auction model seems inferior to the mediator-based model in terms of revenue or
efficiency guarantee due to added capacity. 

In the {\em second scenario}, we initiate a
game theoretic study of current sponsored search auctions, involving {\em incentive
driven} mediators who exploit the fact that these mechanisms are not collusion-resistant.
In particular, we show that advertisers can improve their payoffs by using the services of
the mediator compared to directly participating in the auction, and that the mediator can
also obtain monetary benefit by managing the advertising burden of its advertisers,
without violating incentive constraints from the advertisers who do not use its services.
We also point out that the auctioneer can not do very much via mechanism design to avoid
such for-profit mediation without losing badly in terms of revenue, and therefore, the
mediators are likely to prevail.
Another implication of our results applies to the
traditional media. Publishers of traditional media, such as newspapers and network TV and
radio, have seen significant declines in their audience market shares, as more people have
shifted to the Web as the source for information and entertainment. Their advertising
revenues have decreased significantly as well. Our results show that one way these
traditional media players can retain the loyalty of their advertisers is to manage their
online auctions! By mediating their auctions, they can provide better payoffs to their
clients, and thus prevent them from switching allegiance to the online giants, such as
Google and Yahoo! Other than creating a new revenue source for the traditional media
businesses, it would allow them to retain their own networks and give them precious time
to reposition themselves and figure out the best possible ways to take their content
online, and compete effectively in a new market space.

In conclusion, our analysis indicates that there are
\emph{significant opportunities for diversification} in the internet economy and we should
expect it to continue to develop richer structure,
 with room for different types of agents and mechanisms to coexist.

\subsection*{Acknowledgements}
We thank Paul T. Ryan for insightful comments. The work of S.K.S. was partially supported by his internship at NetSeer Inc.,
11943 Montana Ave. Suite 200, Los Angeles, CA 90049.

\bibliographystyle{abbrv}
\bibliography{ssa}

\begin{thebibliography}{10}

\bibitem{AG07}
Z.~Abrams and A.~Ghosh.
\newblock Auctions with revenue guarantees for sponsored search.
\newblock {\em Third workshop on Sponsored Search Auctions, WWW 2007}.

\bibitem{AGM06}
G.~Aggarwal, A.~Goel, and R.~Motwani.
\newblock Truthful auctions for pricing search keywords.
\newblock In {\em EC '06: Proceedings of the 7th ACM conference on Electronic
  commerce}, pages 1--7, New York, NY, USA, 2006. ACM Press.

\bibitem{med}
I.~Ashlagi, D.~Monderer, and M.~Tennenholtz.
\newblock Mediators in position auctions.
\newblock In {\em EC '07: Proceedings of the 8th ACM conference on Electronic
  commerce}, pages 279--287, New York, NY, USA, 2007. ACM Press.

\bibitem{BCPP06}
T.~Borgers, I.~Cox, M.~Pesendorfer, and V.~Petricek.
\newblock Equilibrium bids in sponsored search auctions: Theory and evidence.
\newblock {\em Technical report, University of Michigan}, 2007.

\bibitem{EOS05}
B.~Edelman, M.~Ostrovsky, and M.~Schwarz.
\newblock Internet advertising and the generalized second-price auction:
  Selling billions of dollars worth of keywords.
\newblock {\em American Economic Review}, 97(1):242--259, March 2007.

\bibitem{Feng-et}
J.~Feng, H.~K. Bhargava, and D.~M. Pennock.
\newblock Implementing sponsored search in web search engines: Computational
  evaluation of alternative mechanisms.
\newblock {\em Informs Journal on Computing}, 2006.

\bibitem{GH03}
A.~V. Goldberg and J.~D. Hartline.
\newblock Competitiveness via consensus.
\newblock In {\em SODA '03: Proceedings of the fourteenth annual ACM-SIAM
  symposium on Discrete algorithms}, pages 215--222, Philadelphia, PA, USA,
  2003. Society for Industrial and Applied Mathematics.

\bibitem{GH05}
A.~V. Goldberg and J.~D. Hartline.
\newblock Collusion-resistant mechanisms for single-parameter agents.
\newblock In {\em SODA '05: Proceedings of the sixteenth annual ACM-SIAM
  symposium on Discrete algorithms}, pages 620--629, Philadelphia, PA, USA,
  2005. Society for Industrial and Applied Mathematics.

\bibitem{Lah06}
S.~Lahaie.
\newblock An analysis of alternative slot auction designs for sponsored search.
\newblock In {\em EC '06: Proceedings of the 7th ACM conference on Electronic
  commerce}, pages 218--227, New York, NY, USA, 2006. ACM Press.

\bibitem{LP07}
S.~Lahaie and D.~M. Pennock.
\newblock Revenue analysis of a family of ranking rules for keyword auctions.
\newblock In {\em EC '07: Proceedings of the 8th ACM conference on Electronic
  commerce}, pages 50--56, New York, NY, USA, 2007. ACM Press.

\bibitem{ORgamebook}
M.~J. Osborne and A.~Rubinstein.
\newblock {\em A Course in Game Theory}.
\newblock {MIT} Press, 1999.

\bibitem{SRGR07}
S.~K. Singh, V.~P. Roychowdhury, H.~Gunadhi, and B.~A. Rezaei.
\newblock Capacity constraints and the inevitability of mediators in adword
  auctions.
\newblock {\em WINE 2007 (to appear)}.

\bibitem{Var06}
H.~Varian.
\newblock Position auctions.
\newblock {\em To appear in International Journal of Industrial Organization}.

\end{thebibliography}

\section*{Appendix}

\paragraph*{ Proof  of  Theorem \ref{rev1}:}
Let
$\eta = \min_{l \leq j \leq K} \frac{\tilde{\gamma}_j -\tilde{\gamma}_{j+1}}{\gamma_j - \gamma_{j+1}}$
then we have
 $\tilde{\gamma}_j -\tilde{\gamma}_{j+1} \geq \eta  (\gamma_j - \gamma_{j+1}) \textrm{ for $ l \leq j \leq K $}$.
At their corresponding minimum {\em SNE} \cite{EOS05,Var06}, the original revenue of the auctioneer without
added capacity and the new revenue of the auctioneer  after adding capacity  are
$R_0 = \sum_{j=1}^K (\gamma_j - \gamma_{j+1}) j s_{j+1}  \textrm{   and }
R = \sum_{j=1}^{\tilde{K}} (\tilde{\gamma}_j - \tilde{\gamma}_{j+1}) j s_{j+1}$
respectively.
\begin{displaymath}
\begin{array}{l}
\therefore R-R_0  = \\
\sum_{j=1}^K \left[ (\tilde{\gamma}_j - \tilde{\gamma}_{j+1}) -  (\gamma_j - \gamma_{j+1}) \right] j s_{j+1} + \\
\\
\sum_{j=K+1}^{\tilde{K}} (\tilde{\gamma}_j - \tilde{\gamma}_{j+1}) j s_{j+1}  \\
\\
       = \sum_{j=l-1}^K \left[  (\tilde{\gamma}_j - \tilde{\gamma}_{j+1}) -  (\gamma_j - \gamma_{j+1}) \right] j s_{j+1} + \\
\\
\sum_{j=K+1}^{\tilde{K}} (\tilde{\gamma_j} - \tilde{\gamma_{j+1})}) j s_{j+1}  \\
\\
     \geq (\gamma_l - \tilde{\gamma}_l)(l-1) s_l +  \sum_{j=l}^K (\eta -1)  (\gamma_j - \gamma_{j+1}) j s_{j+1} \\
\\
+ \sum_{j=K+1}^{\tilde{K}}  (\tilde{\gamma}_j - \tilde{\gamma}_{j+1}) j s_{j+1}  \\
\\
      = (\gamma_l - \tilde{\gamma}_l)(l-1) s_l\\
\\
 + \sum_{j=K+1}^{\tilde{K}}  (\tilde{\gamma}_j - \tilde{\gamma}_{j+1}) j s_{j+1}\\
\\
  -  (1 - \eta) \sum_{j=l}^K (\gamma_j - \gamma_{j+1}) j s_{j+1}\\
\\
\textrm{   and hence follows the theorem.}
\end{array}
\end{displaymath}

\paragraph*{ Missing calculations for Example \ref{example1}:}
\begin{displaymath}
\begin{array}{l}
\textrm{  We have  } \eta = \min_{K \geq j \geq l} \frac{\tilde{\gamma}_j -\tilde{\gamma}_{j+1}}{\gamma_j - \gamma_{j+1}}\\
\\
= \frac{f r^{K-1}(1-r)}{r^{K-1}}=f(1-r). \\
\textrm{ Now, }
(\gamma_l - \tilde{\gamma}_l)(l-1) s_l \\
\\
+ \sum_{j=K+1}^{K+L-1}  (\tilde{\gamma}_j - \tilde{\gamma}_{j+1}) j s_{j+1} \\
\\
\geq (\gamma_K - \tilde{\gamma}_K)(K-1) s_K \\
\\
+ K s_{K+1} (\tilde{\gamma}_{K+1} - \tilde{\gamma}_{K+L}) \\
\\
= r^{K-1}(1-f) (K-1)s_K + fr^KK s_{K+1}. \\
\\
\textrm{Also  } \sum_{j=l}^K (\gamma_j - \gamma_{j+1}) j s_{j+1} = r^{K-1} K s_{K+1}, \\
\\
\therefore  1- \left(\frac{(\gamma_l - \tilde{\gamma}_l)(l-1) s_l + \sum_{j=K+1}^{K+L-1}  (\tilde{\gamma}_j - \tilde{\gamma}_{j+1}) j s_{j+1} }{\sum_{j=l}^K (\gamma_j - \gamma_{j+1}) j s_{j+1}}\right) \\
\\
\leq 1- \frac{r^{K-1}(1-f) (K-1)s_K + fr^KK s_{K+1}}{r^{K-1} K s_{K+1}} \\
\\
= 1- \left(\frac{(K-1)s_K}{K s_{K+1}} (1-f) + fr \right)\\
\\
 = f - fr + (1-f) -\frac{(K-1)s_K}{K s_{K+1}} (1-f) \\
\\
= f(1-r) + (1-f) \left( 1-\frac{(K-1)s_K}{K s_{K+1}} \right) \\
\\
< f(1-r) = \eta.
\end{array}
\end{displaymath}

\paragraph*{Proof  of  Lemma \ref{l1revloss}:}
Let us define
\begin{equation}
i_0 = \max_{1 \leq i \leq K} \left\{i : \gamma_1 f \gamma_1 < \gamma_i \right\} \nonumber
\end{equation}
then $\tilde{\gamma}_j = \gamma_{j+1}$ for all $1 \leq j \leq i_0-1$, $\tilde{\gamma}_{i_0} =\gamma_1 f \gamma_1 $, and
 $\tilde{\gamma}_j \geq \gamma_{j}$ for all $j \geq i_0 +1$. Clearly, $i_0 \geq 1$. Now,
\begin{displaymath}
\begin{array}{l}
R_0 = \sum_{j=1}^K (\gamma_j - \gamma_{j+1}) j s_{j+1} \\
\\
= \gamma_1 s_2 - \sum_{j=2}^K \gamma_j \left[ (j-1) s_j - j s_{j+1}\right] \\
\\
R = \tilde{\gamma}_1 s_2 - \sum_{j=2}^{K+L-1} \tilde{\gamma}_j \left[ (j-1) s_j - j s_{j+1}\right] \\
\\
\therefore R - R_0 = \\
\\
(\tilde{\gamma}_1 -\gamma_1) s_2 - \sum_{j=2}^K (\tilde{\gamma}_j - \gamma_j) \left[ (j-1) s_j - j s_{j+1}\right]\\
\\
    - \sum_{j=K+1}^{K+L-1} \tilde{\gamma}_j \left[ (j-1) s_j - j s_{j+1}\right] \\
\\
\therefore \textrm{    when   }  i_0 \geq 2  \textrm{    we have  } \\
\\
 R - R_0  = -(\gamma_1 -\gamma_2) s_2 \\
\\
+ \sum_{j=2}^{i_0-1} (\gamma_j - \gamma_{j+1}) \left[ (j-1) s_j - j s_{j+1}\right] \\
\\
     + (\gamma_{i_0} - \gamma_1 f \gamma_1) \left[ (i_0-1) s_{i_0} - i_0 s_{i_0+1}\right] \\
\\
     - \sum_{j=i_0+1}^K (\tilde{\gamma}_j - \gamma_j) \left[ (j-1) s_j - j s_{j+1}\right]\\
\\
    - \sum_{j=K+1}^{K+L-1} \tilde{\gamma}_j \left[ (j-1) s_j - j s_{j+1}\right] \\
\\
      \leq - (\gamma_1 -\gamma_2) s_2 + (\gamma_1 -\gamma_2) \left[  s_{2} - i_0 s_{i_0+1}\right]  \\
\\
\textrm{ (recall that $(\gamma_1 -\gamma_2) \geq (\gamma_j - \gamma_{j+1})$ }\\
\textrm{  and $(j-1) s_j - j s_{j+1} \geq 0 $ for all $j \geq 2$ )}\\
\\
       =  - (\gamma_1 -\gamma_2) i_0 s_{i_0+1}\\
\\
      < 0 . \\
\\
\textrm{    when   }  i_0 = 1  \textrm{    we have  } \\
R - R_0  = - (\gamma_1 -\gamma_1 f \gamma_1) s_2 \\
\\
 - \sum_{j=2}^K (\tilde{\gamma}_j - \gamma_j) \left[ (j-1) s_j - j s_{j+1}\right]\\
\\
    - \sum_{j=K+1}^{K+L-1} \tilde{\gamma}_j \left[ (j-1) s_j - j s_{j+1}\right] \\
\\
        < 0.
\end{array}
\end{displaymath}

\paragraph*{Proof  of  Lemma \ref{l2revloss}:}
Let
\begin{equation}
i_0 = \max_{1 \leq i \leq K} \left\{i : \gamma_l f \gamma_1 < \gamma_i \right\} \nonumber
\end{equation}
then $\tilde{\gamma}_j = \gamma_{j}$ for all $1 \leq j \leq l-1$ , $\tilde{\gamma}_j = \gamma_{j+1}$
for all $l \leq j \leq i_0-1$, $\tilde{\gamma}_{i_0} =\gamma_l f \gamma_1 $, and
 $\tilde{\gamma}_j \geq \gamma_{j}$ for all $j \geq i_0 +1$. Clearly, $i_0 \geq l \geq 2$.
\begin{displaymath}
\begin{array}{l}
\therefore R =  \tilde{\gamma}_1 s_2 - \sum_{j=2}^{K+L-1} \tilde{\gamma}_j \left[ (j-1) s_j - j s_{j+1}\right] \\
\\
= \gamma_1  s_2 - \sum_{j=2}^{l-1} \gamma_{j}\left[ (j-1) s_j - j s_{j+1}\right]\\
\\
 -  \sum_{j=l}^{i_0-1} \gamma_{j+1} \left[ (j-1) s_j - j s_{j+1}\right] \\
\\
 - \gamma_l f \gamma_1 \left[ (i_0-1) s_{i_0} - i_0 s_{i_0+1}\right] \\
\\
     - \sum_{j=i_0+1}^{K+L-1} \tilde{\gamma}_j \left[ (j-1) s_j - j s_{j+1}\right]\\
\end{array}
\end{displaymath}
Now let us increase $f$ to $f^{'}$ and denote the new position based CTRs as $\tilde{\gamma_j}^{'}$'s and the new revenue of the auctioneer
as $R^{'}$ then two cases arise - one where $i_0$ does not change and other where it changes to $i_0-1$. \\

Case 1: when $i_0$ does not change by increasing $f$ to $f^{'}$. Clearly, $\tilde{\gamma_j}^{'} \geq \tilde{\gamma_j}$ for all $ j \geq i_0+1$
as we will be choosing elements from a set with larger values. Also recall that $s_i$'s satisfy $(j-1) s_j - j s_{j+1} \geq 0 $ for all $j \geq 2$.
 Therefore, the second last term in the expression of $R$ strictly decrease and
the last term also decreases and we get $R^{'} < R$. \\

Case 2:  When $i_0$ changes to $i_0-1$ by increasing $f$ to $f^{'}$. In this case,
 $\tilde{\gamma}_j^{'} = \gamma_{j}$ for all $1 \leq j \leq l-1$ , $\tilde{\gamma}_j^{'} = \gamma_{j+1}$ for all $l \leq j \leq i_0-2$,
$\tilde{\gamma}_{i_0-1}^{'} =\gamma_l f^{'} \gamma_1 $, and
 $\tilde{\gamma}_j^{'} \geq \tilde{\gamma}_{j}$ for all $j \geq i_0$. Therefore,
\begin{displaymath}
\begin{array}{l}
R^{'} - R = \gamma_{i_0} \left[ (i_0-2) s_{i_0-1} - (i_0-1) s_{i_0}\right]  \\
\\
 - \gamma_l f^{'} \gamma_1 \left[ (i_0-2) s_{i_0-1} - (i_0-1) s_{i_0}\right] \\
\\
 - \sum_{j=i_0}^{K+L-1} ( \tilde{\gamma}_j^{'} - \tilde{\gamma}_j) \left[ (j-1) s_j - j s_{j+1}\right]\\
\\
\leq (\gamma_{i_0} - \gamma_l f^{'} \gamma_1) \left[ (i_0-2) s_{i_0-1} - (i_0-1) s_{i_0}\right] \\
\\
< 0.
\end{array}
\end{displaymath}

\paragraph*{ Proof  of  Theorem \ref{eff1}:}
We have
\begin{displaymath}
\begin{array}{l}
E_0 = \sum_{j=1}^K \gamma_j s_j \\
\\
E= \sum_{j=1}^{K+L-1} \tilde{\gamma}_j s_j \\
\\
= \sum_{j=1}^{l-1} \gamma_j s_j + \sum_{j=l}^{K+L-1} \tilde{\gamma}_j s_j.\\
\\
\therefore
E-E_0 = \sum_{j=l}^K (\tilde{\gamma}_j -\gamma_j) s_j
+ \sum_{j=K+1}^{K+L-1} \tilde{\gamma}_j s_j.
\end{array}
\end{displaymath}
Let
\begin{displaymath}
\beta = \min_{K \geq j \geq l} \frac{\tilde{\gamma}_j}{\gamma_j}
\end{displaymath}
then
\begin{displaymath}
E-E_0 \geq  \sum_{j=K+1}^{K+L-1} \tilde{\gamma}_j s_j - (1-\beta) \sum_{j=l}^K \gamma_j s_j.
\end{displaymath}
and hence follows the theorem.

\paragraph*{ Missing calculations for Example \ref{example2}:}
\begin{displaymath}
\begin{array}{l}
\beta = \frac{\tilde{\gamma}_K}{\gamma_K} = \frac{f r^{K-1}}{r^{K-1}} = f. \\
\\
\textrm{  Also,  }
 \sum_{j=K+1}^{K+L-1} \tilde{\gamma}_j s_j\\
\\
 =  \sum_{j=K+1}^{K+L-1} f r^{j-1} s_j \\
\\
  = fr^K s_K \sum_{j=1}^{L-1} r^{j-1} \alpha^j \\
\\
= \alpha fr^K s_K \left(\frac{1-r^{L-1}\alpha^{L-1}}{1-r\alpha}\right)\\
\\
\therefore 1- \frac{ \sum_{j=K+1}^{K+L-1} \tilde{\gamma}_j s_j}{\sum_{j=l}^K \gamma_j s_j}\\
\\
= 1- \frac{ \alpha fr^K s_K \left(\frac{1-r^{L-1}\alpha^{L-1}}{1-r\alpha}\right)}{r^{K-1} s_K} \\
\\
= 1 - \alpha fr \left(\frac{1-r^{L-1}\alpha^{L-1}}{1-r\alpha}\right) \\
\\
= \frac{1-r\alpha - \alpha r f + f r^{L}\alpha^{L}}{1-r\alpha} \\
\\
= \frac{f(1-r\alpha) + 1 -f - \alpha r + f r^{L}\alpha^{L}}{1-r\alpha} \\
\\
= f + \left( 1 - f \frac{1 - r^{L}\alpha^{L} }{1 -\alpha r} \right)\\
\\
< f   \textrm{  if    } f > \frac{1 -\alpha r}{1 - r^{L}\alpha^{L}}.
\end{array}
\end{displaymath}

\paragraph*{ Proof of Theorem \ref{eff2}:}
Let
\begin{equation}
i_0 = \max_{1 \leq i \leq K} \left\{i : \gamma_l f \gamma_1 < \gamma_i \right\} \nonumber
\end{equation}
then $\tilde{\gamma}_j = \gamma_{j}$ for all $1 \leq j \leq l-1$ , $\tilde{\gamma}_j = \gamma_{j+1}$ for all $l \leq j \leq i_0-1$, $\tilde{\gamma}_{i_0} =\gamma_l f \gamma_1 $, and
 $\tilde{\gamma}_j \geq \gamma_{j}$ for all $j \geq i_0 +1$. Clearly, $i_0 \geq l$.
\begin{displaymath}
\begin{array}{l}
E= \sum_{j=1}^{K+L-1} \tilde{\gamma}_j s_j \\
\\
= \sum_{j=1}^{l-1} \gamma_j s_j + \sum_{j=l}^{i_0-1} \gamma_{j+1} s_j \\
 \\
+ \gamma_l f \gamma_1 s_{i_0} + \sum_{j=i_0+1}^{K+L-1} \tilde{\gamma}_j s_j.\\
\end{array}
\end{displaymath}
Now let us increase $f$ to $f^{'}$ and denote the new position based CTRs as $\tilde{\gamma_j}^{'}$'s and the new efficiency
as $E^{'}$ then two cases arise - one where $i_0$ does not change and other where it changes to $i_0-1$. \\
Case 1: when $i_0$ does not change by increasing $f$ to $f^{'}$. Clearly, $\tilde{\gamma_j}^{'} \geq \tilde{\gamma_j}$ for all $ j \geq i_0+1$
as we will be choosing elements from a set with larger values.
 Therefore, the second last term in the expression of $E$ strictly increase and
the last term also increases and we get $E^{'} > E$. \\
Case 2:  When $i_0$ changes to $i_0-1$ by increasing $f$ to $f^{'}$. In this case,
 $\tilde{\gamma}_j^{'} = \gamma_{j}$ for all $1 \leq j \leq l-1$ , $\tilde{\gamma}_j^{'} = \gamma_{j+1}$ for all $l \leq j \leq i_0-2$,
$\tilde{\gamma}_{i_0-1}^{'} =\gamma_l f^{'} \gamma_1 $, and
 $\tilde{\gamma}_j^{'} \geq \tilde{\gamma}_{j}$ for all $j \geq i_0$. Therefore,
\begin{displaymath}
\begin{array}{l}
E^{'}
= \sum_{j=1}^{l-1} \gamma_j s_j + \sum_{j=l}^{i_0-2} \gamma_{j+1} s_j \\
 \\
+ \gamma_l f^{'} \gamma_1 s_{i_0-1} + \sum_{j=i_0}^{K+L-1} \tilde{\gamma}_j s_j.\\
\\
\geq E + ( \gamma_l f^{'} \gamma_1 - \gamma_{i_0})  s_{i_0-1} \\
\\
> E
\end{array}
\end{displaymath}

\newpage

\begin{table*}
\centering
\begin{tabular}{|c|c|c|c|c|c|c|c|c|c|}
\hline
$i$  & 1 & 2 & 3 & 4 & 5 & 6 & 7 & 8 & 9  \\
\hline
$\gamma_i$ & 1 & 0.6 & 0.5 & 0.4 & 0.3 & 0.2 & 0.15 & 0.10 & 0 \\
\hline
$e_i v_i$ & 26 & 22 & 20 & 18 & 17 & 15 & 12 & 12 & 9 \\
\hline
$r_i=e_ib_i$ & 25 & 20 & 16 & 15 & 14 & 13 & 11 & 10 & 9 \\
\hline
$e_ip_i$ &  20 & 16 & 15 & 14 & 13 & 11 & 10 & 9 & 0 \\
\hline
$r_i^{'}=e_i b_i^{'}$ & 14.2 & 14.2 & 14.2 & 14.2 & 14 & & & &  \\
\hline
$e_i\tilde{p}_i$ & 14.2 & 14.2 & 14.2 & 14 & 13 & & & &\\
\hline
\end{tabular}
\caption{Position based CTRs, true valuations, bid profile, and modified bid profile when $M=\{1,2,3,4,5\}$
(recall that $p_i$ denotes the amount
bidder $i$ pays per-click in the game without mediator and $\tilde{p}_i$ denotes the amount the mediator pays
per-click on her behalf).}
\label{med1}
\end{table*}

\begin{table*}
\centering
\begin{tabular}{|c|c|c|c|c|}
\hline
position $j$ & payoff: & payoff by defecting to $j-1$:  & payoff by defecting to $j+1$: & SNE condition \\
 &  &  &  &   satisfied  \\
&  $u_j=\gamma_j (e_jv_j -r_{j+1})$ &  $u_j^{j-1}=\gamma_{j-1} (e_jv_j -r_j)$ &  $u_j^{j+1}=\gamma_{j+1} (e_jv_j -r_{j+2})$ &(YES/NO) \\
\hline
1 & 1 (26 -20)= & & 0.6 (26-16) = & \\
  &  6          & & 6 & YES \\
\hline
2 & 0.6 (22 -16)= & 1(22-20)= & 0.5 (22-15) = & \\
  &  3.6          & 2         & 3.5  & YES \\
\hline
3 & 0.5 (20 -15)= & 0.6 (20-16)= & 0.4 (20-14) = & \\
  &  2.5         & 2.4        & 2.4  & YES \\
\hline
4 & 0.4 (18 -14)= & 0.5 (18 -15)= & 0.3 (18-13)=& \\
  &  1.6         & 1.5 & 1.5 & YES \\
\hline
5 & 0.3 (17-13)= & 0.4 (17-14) = & 0.2(17-11)= & \\
  &  1.2         & 1.2 & 1.2 & YES \\
\hline
6 & 0.2 (15-11)= & 0.3 (15-13)= & 0.15(15-10)= & \\
  & 0.8          & 0.6          & 0.75 & YES \\
\hline
7 & 0.15 (12-10)= & 0.2 (12-11)= & 0.10 (12-9)= & \\
  & 0.3           & 0.2          & 0.3 & YES\\
\hline
8 & 0.10 (12-9)= & 0.15 (12-10)= & 0 (12-0) = & \\
  &   0.3        & 0.3           & 0 & YES \\
\hline
9 & 0            & 0.10 (9-9)= 0 & 0 & YES \\
\hline
\end{tabular}
\caption{Verifying the SNE conditions}
\label{med2}
\end{table*}

\begin{table*}
\centering
\begin{tabular}{|ccc|c|c|c|c|c|}
\hline
& $j$ & & $x_j$ &  $r^{*}$ & $r$ & $l$ & mediator's payoff $U_M$: \\
&  & & $=e_jv_j - u_j$ &   & &  &$ \alpha \sum_{j=1}^{l-1} ( r_{j+1} - r) \gamma_j$ \\
\hline
& 6 &  & 15-0.8 = & & & & \\
& &    &  14.2    & & & &\\

& 7 &   & 12-0.3=  & 14.2 & 14.2 & 4 &  7.28 $\alpha$\\
& &    & 11.7     &      &  &   &  \\

& 8 &   & 12-0.3 = & & & & \\
& &    &  11.7   & & & & \\

& 9  &  & 9-0 = & & &  &\\
  & &  &  9    & & & & \\
\hline
\end{tabular}
\caption{Computing $r$ and $l$: $x_j:=(1 - \frac{\gamma_j}{\gamma_1}) e_j v_j + \frac{\gamma_j}{\gamma_1} r_{j+1}=
e_jv_j-\gamma_j(e_jv_j-r_{j+1})=e_jv_j-u_j$ (as $\gamma_1 =1)$ }
\label{med3}
\end{table*}

\begin{table*}
\centering
\begin{tabular}{|c|c|ccccc|c|c|c|}
\hline
position $j$  & 1 & 2 & 3 & 4 & 5 & 6 & 7 & 8 & 9 \\
\hline
bidder $i$ assinged to $j$ & 6 &  1 & 2 & 3 & 4 & 5 & 7 & 8 & 9 \\
\hline
$e_i v_i$ & 15 & 26 & 22 & 20 & 18 & 17  & 12 & 12 & 9 \\
\hline
$r_i=e_ib_i$ &13 &  25 & 20 & 16 & 15 & 14 & 11 & 10 & 9 \\
\hline
$e_i p_i$ & 11 & 20 & 16 & 15 & 14 & 13 & 10 & 9 & 0 \\
\hline
$r_i^{'}=e_i b_i^{'}$ &  & 12 & 12 & 12 & 12 &12 & & &  \\
\hline
$e_i \tilde{p}_i$ &  & 12 & 12 & 12 & 12 & 11& & & \\
\hline
mediator's payoff $U_M$: & & & & 22.8 $\alpha$ & & & & & \\
\hline
\end{tabular}
\caption{Sliding positions could improve mediator's payoff}
\label{med4}
\end{table*}

\end{document}